\documentclass[12pt]{article}
\usepackage{epsfig}
\usepackage{graphicx}
\usepackage{fullpage}
\usepackage{setspace}
\usepackage{amsmath}

\def\tstrut{\vrule height 1.2em depth 0.5em width 0pt}

\begin{document}
\doublespacing

\vskip1.5cm
\begin{center}
{\Large\bf  Quirks and strings attached as the ultimate communication and acceleration devices}
\tstrut
\\{\bf Shmuel Nussinov}\footnote{\tt nussinov@post.tau.ac.il}
\end{center}

\vskip1cm
\begin{center}
Raymond and Beverly Sackler School of Physics and Astronomy\\
Tel Aviv University, Tel Aviv, Israel.\\
\end{center}
\begin{abstract}
 We point out that if a certain variant of ``Quirks", particles that carry ordinary color and some other color' exist, then we can have a completely novel and efficient mode of long distance communications and of acceleration to very high energies. For very low scale $\Lambda'$ the scale of the new gauge group in the theory, and associated string tension of the new color'  the Quirks can be captured in ordinary materials. Having then the Quirk Q' and anti-Quirk $\bar{Q}'$  in two separate piezoelectric crystals at arbitrarily far out points A and B allows Alice and Bob at these locations to communicate by generating transverse waves along the connecting color' string. Also releasing the Quirks allows them to collide at extremely high energies.     
\end{abstract}

\newpage
\section{Introduction}
All known elementary particles fit in the framework of the standard model(SM). Hopefully the LHC (Large Hadron Collider) will reveal new physics and new particles associated with spontaneous breaking of the $SU(2)_L\times U(1)$ of the standard model. At present after the runs at $7$ and $8$ TeV there is no evidence for such new physics and the only particle discovered - the scalar at $126 GeV$ - looks very much like the SM, Brout-Englert-Higgs, boson.  This suggests paying more attention to the possible discovery of \textit{unexpected} new particles and interactions. Such discoveries may broaden our prospective and provide deeper insight. Here we point out that a particular new such particle may have fantastic technological implications as well.

In reversing the question ``Who ordered it?" asked by I.I. Rabi after the muon was discovered, which with the masses and mixing of the other leptons the quarks still awaits deeper explanation, let us ask: What features should a new particle have if it were \textit{technologically} useful? (so that ordering it is worthwhile).

 An obvious requirement is longevity. Indeed all particles discovered over the last 50 years are short lived. This is particularly unfortunate in the case of the muon. Had it been sufficiently long lived, it could be used to catalyze fusion. A muon replacing one of the two electrons in a molecule of DT pulls the nuclei to distances $200=m(\mu)/{m(e)}$ times smaller than the usual separation of $1/{(m(e)\alpha)}= r_{Bohr}\approx 0.5 10^{-8} cm$. This reduces the tunneling barrier between the nuclei and dramatically accelerates their fusion. However in some $D+T \rightarrow {He+n}$ reactions the muon remains bound to the newly formed $\alpha$. This and the short, $2.2$ microsecond, lifetime reduces the efficiency of muon catalyzed fusion to bellow the break-even point.~\footnote{catalyzed fusion was reconsidered when the possibility that dark matter consists of (positively and negatively) charged massive particles (?Champs?) was suggested by De-Rujula, Glashow and Sarid\cite{de -Rujula} }

Another requirement is that it be produced in accelerators, by cosmic rays or be otherwise available. Future technologies may allow reaching energies far higher than the $\sim 10^4 GeV$ of the LHC. Thus a second requirement is that the new particle X be relatively light $2M(X)\le f{E_{max}}$ where $E_{max}$ is the maximal energy achievable at any given epoch (the factor $2$ is there because the $X$ particles are usually pair produced and the extra 'fudge ' factor  $f\le{1}$ is there because for proton colliders the p.d.f.'s, the parton distribution functions reduce the actual energy available in say gluon-gluon collisions. Using what is available now the bound is  $M(X)\le{(1-3)} TeV$.

Production in hadron machines such as the LHC, is optimized if the new $X$ particle carries ordinary QCD color though just as in the case of the Higgs particle it could also be produced weakly via vector boson fusion- albeit with a smaller cross-section. 

Finally we should be able to control and manipulate the particle which means that it should couple to ordinary matter and/or electromagnetism.

Clearly the above are necessary conditions. Since only a handful of the new particles are usually produced - far smaller than the typical Avogadro number of atoms or electrons in ordinary macroscopic material samples, there may be too few $X$ particles to allow practical implications.

 In addition we need to satisfy many particle and astro- particle physics constraints. The physics framework in which the $X$ particle is introduced should not break exact gauge symmetries and should be part of a consistent theory with no  "Anomalies" of various types. Differently stated the theory should have a good ' Ultra-violet" completion- which is in particular the case for asymptotically free theories.

 Many other constraints follow from the demand that the new $X$ particle and/or additional particles associated with it will not spoil the successful predictions of the standard cosmological scenario: $\bar{X}$ and $X$ should annihilate efficiently and early enough so as not to exceed the fraction $\Omega(CDM) \sim {0.2}$ of the critical energy density and not distort the precise Planck CMB Cosmic Microwave background spectrum. Also if other light particles come along with $X$, they should contribute at most the equivalent of half a light neutrino at the time of big bang nucleo synthesis and if these light particles are stable remnants then the latter should not lead to over-closure or unacceptable forms of dark matter. 

Finally we need to verify that particles produced in high energy cosmic ray events do not leave observable relics which should have already be seen.

The following in not a systematic search for useful particles in the space of all reasonable beyond the standard model physics scenarios. Rather we note that particles which may fit all the above requirements have already been suggested. Over the years various existing or hypothesized particles with potential technological use have been encountered. Neutrinos, in particular intense high energy neutrino beams with large penetration lengths were suggested \cite{Charpack} for whole earth tomography or more local yet deep geological surveying \cite{Loewy}. The magnetic monopoles are time honored stable theoretically highly motivated particles that could indeed have far reaching applications. However the likely high($\sim10^{17}GeV$) mass of such particles in a GUTs (Grand Unified Theories) context and the exponentially suppressed cross-section for their production in colliders \cite{Drukier} are severe obstacles. Small, rotating black holes of masses $\ge 10^{15}gr$ are stable against Hawking radiation and if towed to given controlled location could be a depository of dangerous waste  and a clean source of energy. Despite their tiny Fermi size these are not regular elementary particles and certainly cannot be produced at the LHC.

``Quirk" -denoted here as $Q's$-were introduced by Luthi $\cite{Kang1,Kang2}$. Our study $\cite{Jacobi}\cite{Nussinov1}$ suggested that a subset of Quirk models with properly chosen one parameter all the above constraints can be satisfied. The Quirks were introduced as " why not?" particles, not motivated by any theoretical arguments, but not ruled out either. For sub $TeV$ masses the Quirks can be produced at LHC with reasonable rates and have dramatic signatures. However the LHC triggers optimized to search other more motivated particles such as super-symmetric partners, may miss these signatures.

The new observation that is our focus here is that if the scale of the new "Quirky" color is low enough,$\le{100 eV}$ which may be required by other considerations, then even few Quirks can be useful. When captured inside appropriate grains the strings connecting them may provide direct point to point long distance contact with no $1/{R^2}$ fall-off with distance which limits all presently known forms of communication.

\begin{figure}
\begin{center}
\includegraphics{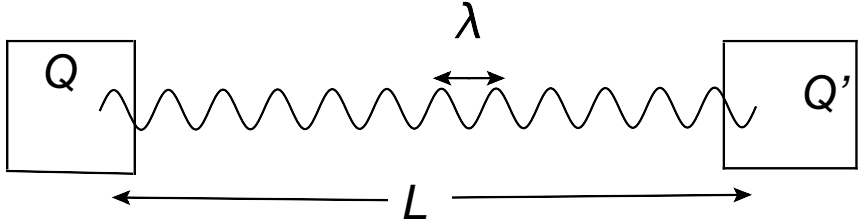}
\caption{A schematic description of how one can send wave packets of transverse phonons of wave length $\lambda$ along extended Quirky strings in order to communicate with a far away  party at a distance $L$.}
\end{center}
\end{figure}

In the next section, we briefly review Quirks and their particle- and astro- particle phenomenology. In particular we discuss at some length the unique features of the residual light glue-balls gb's  associated with the low scale new gauge theory.

In Sec III we indicate how one might use the Quirks and the color' string connecting them for a novel mode of point to point unattenuated communication and discuss possible sources string damage and of noise. For the $SU(3')$ variant of the Quirk model, string networks for many parties communications may be possible.

In Sec IV we elaborate another possibility of accelerating the Quirks and matter attached to it to extremely high energies. 

Sec V presents some further discussion of Quirks at the LHC. 

In Sec VI we discuss shortly models for the Quirk mass and the ultra dense matter that Quirk models afford if we have a Quirk asymmetry 

Few final comments are presented in Sec VII. 

\section {The Quirks-some generalities and viability}
Some of the material in this section was discussed in the above four references. We present the arguments limiting the range of allowed scale $\Lambda'$ of the color' and discuss key issues. We elaborate on the new background due to glue-balls made of the gluon's of the new gauge group $SU(N')$ and present various estimates of its density at present. 

The Quirk model adds to the $SU(2)_LXU(1) XSU(3)_{Color}$ of the standard model (SM) another gauge group $SU(N')$ with $N'=2$ or $3$.(Higher $N'$ values are excluded by the effect of the light unconfined gluon's on nucleo synthesis and N'=2 with three rather than eight gluon's is preferred). The Quirks  transform as $(3,N')$ under the ordinary and new color' and anomalies are avoided by having them  $SU(2)_LXU(1)$ singlets. 

The Quirks should have a mass $M(Q')\ge{1/2 TeV}$ as otherwise they would have been discovered at the LHC. Carrying \textit{two} conserved color charges the $3N'$ Quirks are absolutely stable. The remaining parameter of the model is the scale $\Lambda'$  of the new $QCD'$ or the mass of the lightest scalar, $g'g'$ glue'-ball $m(g'g')\sim {(7-10)\Lambda'}$ for $N'=3$. \footnote{Glueball masses were calculated in QCD for $N_{c}=3$ in the quenched approximation where one neglects the effect of the quarks, are $m(gg)\sim {7\Lambda(QCD)}$. Early extensive calculations were done by Don Weingarten and collaborators and for a more up-to-date work see \cite{Morningstar}. The quenched approximation is fully justified in the present case with very heavy Quirks}   

Apriori there is no connection between the scale of a confining theory and the masses of the particles in its fundamental representation. Thus the mass of the top quark is almost $10^3$ times $\Lambda(QCD)$. The scale at which the running coupling of the gauge theory becomes strong $\sim{1}$ has the form $\Lambda\sim {\exp{-b_0/{g^2}}}$ with $g$ a coupling at some high scale and $b_0$ is the first term in the weak coupling expansion of the $\beta$ function of the theory. If we start $SU'(N'=3)$ with a weaker coupling $g'$ at the high scale ,or use the same coupling but choose $N'=2$, then much longer running is required for $g'$ to become strong and $\Lambda'$ can be tiny. 

In the following we assume that $\Lambda'$ is indeed tiny:
\begin{equation} 
\Lambda'= 1-100 eV \textrm{, } m(gb')=7-700 eV
\end{equation}
 $\Lambda'$ is $\sim{10^{11}}$ times smaller than the the Quirk mass. While appearing artificial this does not require fine tuning. We do however need a mechanism for generating the large Quirk mass an issue which we will briefly address In Sec V.  

 The rate of quirk production at LHC is $N'$ times that of a heavy quark of the same mass. The QCD cross-section was calculated for cm energy $W=13 TeV$ to several loops \cite{Mitov}  It varies between $10^{-35}cm^{2}$ and $10^{-37}cm^{2}$ for $M(X)=(1-3) TeV$ respectively. \footnote{Because of interest in SUSY partners extensive calculations of the production of squarks of various masses and at many energies have been done in great detail. Quirks cross-sections can then be readily inferred by multiplying by $2N'$ to account for the effect of spin and the extra multiplicity due to color'. }
 
In $10$ years of operation with integrated luminosities of $10^{42}cm^{-2}$ the LHC will collect $10^{17}$ p-p collisions at center mass energy of $W=13 Tev$. This is roughly the number of interactions at energies $\ge{10^{17}eV}$ (the fixed target equivalent of 13 TeV center mass energy) of cosmic rays (C.R s) falling on a square kilometer during a year, making the LHC a better place to look for Quirks. For $M(Q')\ge{10 TeV}$ future accelerators at $10$ times enter mass energies or $100$ times higher C.R equivalent energies will be relatively better yet. This is so in particular if, as the AUGER experiment suggests, the UHE CR's are Iron nuclei with only $1/{56}$of the energy per nucleon. That the collisions at LHC happen at beam intersections inside the (very!) well instrumented CMS and ATLAS detectors is clearly another very important factor. 

An important advantage for the technological application which require finding and picking up the Quirks is the shorter range of the stopping quirks from the LHC as compared with the $\sim{10^3}$ times longer penetration depths into earth of the CR produced Quirks of $10^3-10^4$ higher laboratory energies.  
 
During earth's lifetime a sufficient number of such heavy particles can be produced by UHE C.R. If they bind to nuclei they would manifest via a tiny admixture of anomalously heavy isotopes. Making sure that this does not exclude various quirk models and those that we will be interested in particular is one of the many points that have to be addressed.

 \subsection{Subsection (ii)} 
An important observation \cite {Jacobi} is that both $Q'$s and $\bar{Q}$'s can bind to medium-heavy nuclei via the following steps:

 Step a)The $\bar{Q'}$ and $Q'$ produced at the LHC or at the top of the atmosphere by cosmic rays, have relative momenta $P$ and kinetic energy in the center mass Lorentz frame of the Quirks:$K.E=P^2/{2M(Q')}$ which are $\sim{(0.5-0.1)}M(Q')$. QCD and $SU(N')$ flux-tubes/ strings connecting them eventually form.

Step(b) After only a $\sim{1-2 Fermi}$ separation a $\bar{u}u$, $\bar{d}d$ and in $1/{10}$ of the cases a $\bar{s}s$ quark pairs are generated breaking the QCD string. The $Q'$s picks up the $\bar{q}$ forming a supper-heavy ?meson?: $\bar{M}= Q'\bar{q}$ and the  $\bar{Q}'$ binds the $q$ making the conjugate meson $M=\bar{Q'}q$. We use $q$ for light up $u$, down $d$ and strange quark $s$ flavors. 

The single most important feature and a key to \textit{all} the following is that the remaining $SU(N')$ string between the $\bar{Q'}$ and $Q'$ never breaks. This is so because the lightest particles  carrying color' are the Quirks with a large mass:
$M(Q')\sim {10^{11}\Lambda'}$. String breaking requires creating a $Q'\bar{Q'}$ pair forming the two ends of the two new strings. Even for long strings with total energy exceeding $2M(Q')$ this pair creation is impossible. Local color conservation implies that the creation of $Q'\bar{Q'}$ happens locally and only an energy $\sim{\Lambda'}$ is locally stored a string bit of natural length: $a'={\Lambda'^{-1}}$.

Indeed Schwinger's formula for producing electron-positron pairs in a uniform E field \cite{Schwinger} adapted to the Chromo- electric flux tube in QCD \cite{Casher1} yields a truly tiny rate:
$Rate{(\bar{Q'}Q')}\sim \exp{(-M(Q')^2/{\Lambda'^2})}$ or a rate of $\exp{(-10^{22})}$ for pair creation and string breaking. 

The low tension $\sigma\sim{4\Lambda'^2}$, $SU(N')$ string plays a key role in enhancing Quirk annihilation thereby avoiding difficulties with excessive residual Quirks. It is also the main ingredient into our proposed novel technological applications.
This string does not however manifest before the $Q'$ and $\bar{Q'}$ separate by a distance $d_{T.P}\sim(K.E./\sigma')$ where the string tension is $\sigma'\sim{4\Lambda'^2}$ and $d_{T.P}$ is the distance to the turning point at which the initial kinetic energy $K.E$ of the Quirks is transformed into the energy of the extended string, the Quirks stop, start moving towards each other etc. For $\Lambda'=(100-1)eV$ and kinetic energy of the Quirks in their cm (center mass) Lorentz frame  between $0.1$ and $0.3 TeV$ this $d_{T.P}$ varies between $3.10^2$  and $10^7$ cm with the shorter $d_{T.P}$ corresponding to  the higher $\Lambda'$ values.
\begin{equation}
\begin{split}
\label{Distance to turning point}
&\textrm{Distance to turning point}\\
&d_{T.P}= (300-10^7) cm , for\ \Lambda'=(100-1)eV, and K.E = 0.1-1 GeV
\end{split}
\end{equation}

Once the distance $r$ between $\bar{Q'}Q'$ becomes smaller than ${a'= 1/{Lambda'}}$ we have a Coulomb like $-\alpha '/r$ attraction between $Q'$ and $\bar{Q'}$.
 
\subsection{Subsection (iii)}

 The subsequent evolution of the $\bar{M'}M'$ system is very different depending on where the initial $\bar{Q'}Q'$ production event took place -in vacuum via the collision of an UHE CR with the interstellar medium, in the atmosphere or in LHC detectors.

(a)In vacuum the conserved low $l=0,1$ relative (cm) angular momentum of the formed $\bar{Q'}Q'$ forces them or the corresponding $M'\bar{M'}$ mesons to retrace their paths and re-collide. Initially the cross-section for $\bar Q'-Q'$ annihilation or for rearranging into $\bar{Q'}Q' $ ?Quirkonium? and a $\bar{q}q$ is quite small and one needs to first emit the extra energy via gluons (or gluon's). However in each collision the light quarks inside $\bar{M'} M'$ produce pions carrying typically an energy of
$0.5 GeV$. Thus a fraction 
$0.5 GeV/{2M(Q')}\sim 10^{-3}-10^{-4}$ of the initial cm kinetic energy $K.E=0.1-1 TeV $ is emitted in each collision.
 In $10^3-10^4$ collisions the $M'$ and $\bar{M'}$ slow down enough so as to facilitate the rearrangement reaction into Quirkonium and an ordinary $\bar{q}q$ meson. A quick cascading between the Quirkonium levels then follows, where in  each step a $g'$ (still mass-less and unconfined at Quirkonium size scales!) is emitted. Eventually the ground-state is reached where the $Q'\bar{Q'}$ efficiently annihilate.

The quirks move with a velocity which for most of the oscillation period is a size-able fraction of the speed of light. The time for traveling the total distance of 
$(10^3-10^4)d_{T.P}$ required for the slowing down and Quirkonium formation  is then:
$t_{Q'.F}=(10^{-7}-10^{-1})Sec$. This time is in the $\bar{Q'}Q'$ system which is close to the $(p-p)$ proton -proton collision or the lab frame for the LHC. For fixed target production in cosmic ray collisions this time is prolonged by the Lorentz factor $\gamma=M(Q')/m(N)=10^3-10^4$ so as to become:
\begin{equation}  
t_{Q'.F}(in\ UHE\ CR\ production)=10^{-1}- 3.10^{4} sec 
\end{equation}
\cite{equation time for Q.F in CR collisions}
where the lower and higher times are relevant to the case of $\Lambda'=100 eV$ and $\Lambda'=1 eV$ respectively.
\{foot-note {We discussed at length \cite {Jacoby}the shortening of the $SU(N')$ string leading to $\bar{Q'}Q'$
annihilation in the early universe after the $SU(N')$ confining phase transition. The $Q'$s are much slower than the case discussed above and start at much shorter relative $\bar{Q'}Q$' separation. Also the frictional drag due to interactions with the $g'g'$, the glue-balls of $SU(N')$ which we denote as $gb's$ is important. We found that this latest third stage of annihilation does indeed reduce the relic $Q'$ densities to acceptable levels}

(b)When produced at the top of the atmosphere or at the LHC The Quirks interact with the air, with matter inside the detector and eventually with water or Rock materials. The $d(E)/{dx}$ energy loss due to ionization by the $1/3$ and $2/3$ charged $M_d$ and $M_u$ and due to nuclear interactions are small as compared with the initial kinetic energy (K.E) of the Quirks in their Center of mass Lorentz frame. Thus the $Q'$ and $\bar{Q}'$ produced at LHC or in the atmosphere are not likely to stop before hitting the rocks or earth's surface  Using $1-3 gr/{cm^3}$ densities for the ocean and earth's crust and estimating the $M'-N$ (N=nucleon) cross-sections by quark counting to be $\sigma\ (M'-Nucleon)\sim (1/2) \sigma(\pi-Nucleon)=10 mb$ we find that the $Q'$ stops after $100 meter -100 Km$ for the case of LHC and Cosmic ray produced Quirks in stopping time of $t_{stop}=10^{-6}-10^{-3}Sec$ respectively.

 We note that the charges at the two ends and the corresponding  energy losses  $dE/{dx}$ via ionization become different after charge exchange interactions. Also the nuclear interactions of the anti-quark $\bar{q}$ and hence of the $\bar{M'}$ heavy meson with nucleons are stronger than those of a quark $q$ or the $M'$ meson. The corresponding different rates of slowing down then increases the separation between the $M'$ and $\bar{M'}$ at ends of the string when the latter eventually stop.

The $\sim (3-4) MeV\ d-u$ mass difference implies a similar difference between $M_d$ and $M_u$. The rate for the $\beta$ decay $\Gamma(M_d\rightarrow{M_u})$ is $G_F^2\Delta^5/{192\pi^3}$. It yields a decay time of $\sim{10-100 Sec}$ which is far longer than the stopping time of the Quirks produced at the LHC or by UHE CR's . This implies that we have equal number of stopped $M_d$ as $M_u$.~\footnote{In the ten percent of the cases where $\bar{s}-s$ quarks have been initially produced by breaking of the QCD string we have $M_s$ ( rather than $M_d$ or $M_u$) mesons and the faster $\beta$ decay of the s quark may allow it to decay into u quark before stopping} 

 \subsection{Subsection(iv)}

It is well known that $\bar{K}$ mesons containing a $\bar{u}$ or a $\bar{d}$ anti-quark are attracted to nucleons and to nuclei and form tightly bound Hyper-nuclei. The same holds here for the heavier $\bar{M}=Q'\bar{q}$.

For $\bar{Q'}q$ mesons the situation is not as clear cut. For the lightest in the heavy meson series, namely the Kaon $K$ ,the  $KN$ scattering length is repulsive. This is due to the fact that in integrating the $KN$ potential over the extended $KN$ state, the short range repulsive interaction contributes more than the attractive negative part of the potential. The repulsion attributed to $\omega$ vector meson exchange, reflects the Pauli exclusion between the quark in the mesons $M'$ and those inside the nucleon. The longer range force between $M'$ and nucleons is due to the $2\pi$ exchange 
\footnote{single $\pi$ exchange is forbidden by parity}. When coupled to a $0^+$ state in the $t$ channel this generates ,as in scalar gravity, attraction between the meson $M'$ and all nucleons. The $0.6 Fermi$ range of the attractive force is longer than that of the  $\sim{0.3} Fermi$ repulsion. The reduced mass of the heavy meson- nucleus $M'-(A,Z)$ system is $2A$ times larger than for $KN$. This then may allow localizing, with minimal cost in kinetic energy, the meson $M$ at some distance $R^*$ from the surface of the $(A,Z)$ nucleus where it can still benefit from the attractive interaction and experience less of the  repulsive shorter range interaction. This attraction is largely diluted by not having a sharp nuclear surface but rather a gradually decreasing density over a distance of $\sim{2 Fermi}$. Still our estimates $\cite{Z. Nussinov}$ suggest that this yields a bound state - albeit with much weaker binding than that of an $\bar{M}$.
Furthermore an $M_d$ with charge $-1/3$ localized at a distance of $R^*= 6 Fermi$ from the center of a nucleus $(A,Z)$ has extra Coulomb binding of $\delta(B)_{Coulomb}=Z\alpha(em)/{3R^*}$ which for $Z=50$ can be as large as $4.5 MeV$. This helps binding and prevents the decay of the bound $M_d$ to a $M_u$ as the latter would have twice as large, $10MeV$, repulsive interaction.

\subsection{Subsection( vi)}
the nuclei to which the $Q'\bar{q}$ and $\bar{Q}'q$ bind are part of atoms and in rock, (but not in ocean water!), the atoms can be part of a lattice. The binding of the high $Z$ atom to it's equilibrium location there can be $100 eV$ or more. For a $2 Angstrom $ lattice constant only force exceeding: 
\begin{equation}
F_{crit}= 100 eV/{2Angstrom}=10^5 eV^2
\end{equation}
can pull the atom and the Q' in it out of the lattice. For 
\begin{equation}
 Lambda'\le{100 eV}
\end{equation}
and string tension:
\begin{equation}
\sigma'\sim 4\Lambda'^2\le{4.10^4 eV}  
\end{equation}
The constraint is satisfied.
 We will consider the range $100 eV\ge{\Lambda'}\ge{1 eV}$ and  quote results for various choices of $\Lambda'$ in this range. In particular we find for $\Lambda'=1-100 eV$,
\begin{equation}
\sigma'=4\Lambda'^2=4eV^2-40.10^3 eV^2 , m(gb')=7 e.V-700 eV
\end{equation}
respectively.

 \subsection{Subsection (vii)}

 we next turn to the various constraints that the Quirk model should satisfy and verify that they do indeed hold:

a) The mew $SU(N')$ is asymptotically free. Also adding $N'=2-3$  color triplet Quirks does not modify for $M(Q')\ge{TeV}$ the observed running of $\alpha(QCD)$ and keeps QCD asymptotically free.
 
 b) Too many relic quirks remain after the early annihilation into $gg$ or $g'g'$ pairs at freeze-out temperatures $T_{f.o}= M(X)/x$ with $x\sim {20}$ and we need to verify that their abundance is further reduced. Indeed after color confinement at $T_{con}\sim{\Lambda(QCD)= 200 MeV}$ further annihilation can  follow the $Q'\bar{q}+\bar{Q'}q \rightarrow \bar{Q'}Q'+\bar{q}q$ rearrangements. These indeed dramatically reduce the co-moving Quirk number densities so that any later annihilation after $t\sim{Sec} $ or temperature $T\le{MeV}$ cannot modify the successful standard big bang prediction for the abundance of light nuclei.~\footnote{The importance of the efficient second stage of annihilation in reducing the numbers of heavy colored particles was recognized earlier in the context of split SUSY and potential difficulties which might have otherwise arise due to a relatively long lived gluino \cite{Arkani-Hamed}}

 Much later when the temperature of the gluon' $g'$ gas is:
\begin{equation}
 T'=T'_{con}\sim\Lambda'{\sim {1-100 eV}} 
\end{equation}

The Coulomb attraction between $\bar{Q'}$ and $Q'$ due to $g'$ exchange crosses over to the constant pull of the $SU(N')$ string and the remaining $Q'$ and $\bar{Q'}s$ or the $\bar{M'}$ and $M'$ mesons get confined on $a'=\Lambda'^{-1}$ scales.

As shown in some detail in \cite{Jacoby}, this eventually brings the $Q'$ and $\bar{Q'}$ which did not bind to heavy nuclei close together so that they efficiently annihilate. The net result is to reduce the number of surviving Q's and satisfy all limits.

For $SU(2)'$ we could have in addition to the $Q'\bar{Q'}$ mesons $Q'Q'$ bosonic baryons of the same mass. However since at short distances the Coulomb color interactions are much stronger than those of color' these di-Quirks are in $l=0,$ in relative s-wave, wave function forming the color anti-symmetric combination which is a $\bar{3}$ of ordinary $SU(3)_c$. Also the one gluon exchange QCD interactions is more attractive when in ordinary spin space the two quirks are in the symmetric $S=1$ state. To ensure overall anti-symmetry the Quirks must then be symmetric also in the internal $SU(2')$ color and therefore be triplet of  of color'.  These will form  ?Quirky ?  $Q'Q'q$ heavy baryons $B'$ which are color singlets but carry the fractional charges of the quarks and are $SU(2')$ triplets Eventually when $SU(2')$ confines these will pick up a g' the triplet gluon' to make also color' singlets. Too many relics of this type are avoided by the rearrangement processes:

$Q'Q'q+\bar{Q'}\bar{Q'}\bar{q}\rightarrow{2\bar{Q'}Q'+\bar{q}q}$.

For $N'=3$ We can form, albeit more rarely, analog of QCD baryons by combining three $M'$s. With the strong Coulomb color binding of the heavy Quirks in $Q'^3$ baryon's favoring the rearrangement
$(Q'\bar{q})^3\rightarrow{Q'^3+ \bar{q}^3}$ .

The Quirky $Q'^3$ baryon's are charge neutral. The most stable such baryons should have symmetric spatial Quirk wave function. For the $Q'^3$ wave function be completely anti-symmetric It should be completely symmetric in the $3'$ indexes, forming the $10'$ color' representation where the tighter ordinary color binding overcomes the weaker color' repulsion. Also the would be color' strings connecting such baryons at long distances can be screened by $g'g'$ pairs. Finally for very massive $Q'$ the $Q'^3$ baryons are almost point like color neutral object largely suppressing the residual nuclear interactions. The baryon's need not then form heavy isotopes whose abundance is experimentally limited and their number density is small enough so that they do not constitute dark matter. This indeed should be the case as even tiny $1/{M(Q')^2}$ $Q'-N$ ( N=Nucleon)  cross-sections would have then been detected in  the searches for cold DM at Exon and Lux.
 
c)The WMAP/PLANCK bound stating that at nucleo-synthesis there were at most the equivalent of half a neutrino light degrees of freedom is (barely!) satisfied for the eight gluon's in $N'=3$ $SU(N')$ thanks to the extra entropy pumped into the photon background but not into the cooler $g'$ CMB' radiation by the eventual annihilation of all SM particles . \footnote{Quirk annihilate much more to the stronger coupled standard model $gg$ gluons than to to $g'g'$ pairs.}

Clearly $SU(N'=2)$ with only three gluon's would be preferred from this point of view- and also by the fact that the smaller $b_0$ coefficient in the $\beta$ function of the $SU(2')$ gauge theory naturally yields a very small $\Lambda'$.

d) Stable particles of mass $M(X)\gg{{m(N)}}$ ($N$ stands for nucleon and $m(N)\sim{GeV}$)with negative charges and or strong interactions produced by the UHE CRs (cosmic rays) can attach to nuclei in ocean water and form ultra-heavy isotopes on which there are strong experimental upper bounds.\cite{Themick} The bounds are most stringent for hydrogen like heavy Isotopes with unusual, small, charge to mass ratio. A serious problem which Quirk model builders have to address is then the cumulative effect of Quirk  production by CR's and the resulting ultra-heavy isotopes.

For the case of a $\bar{M'}= Q'\bar{u}$ meson the charge of the $\bar{M'}p$ (p=proton) tightly bound system is only 1/3e and the charge to mass ratio is correspondingly smaller and the unusual isotope even more striking.  However in water, unlike the case of Quirks firmly lodged inside rock, the string tension $\sim{4\Lambda'^2}$ can pull together the $\bar{Q'}$ and $Q'$ at the  ends of the string. Once the distance between $\bar{Q'}Q'$  becomes smaller than $a'= 1/{\Lambda'}$ we have a Coulomb force between $Q'$ and $\bar{Q'}$. The attractive $g'$ exchange interaction $\alpha'(r)/r$ overcomes ordinary electric Coulomb $\alpha(em)Z/{3r}$ so long as $3\alpha'\ge{Z\alpha(em)=Z/{137}}$ and $\alpha'(a')\sim(1)$ clearly satisfies the above condition.

Furthermore the very peculiar force pattern acting on quirks or the $M'$ or $\bar{M'}$ mesons and/or nuclei to which they are attached which always includes the extra pull of the $SU(N')$ string may disable standard search and separation methodologies.

Thus, if one string end is attached say to a $\bar{Q'}$  under the ocean floor, boiling the ocean water in which the $Q'$ say, resides , will not evaporate it due to the much stronger pull of the string. Searching for anomalous isotopes by repeated boiling and condensation followed by  mass spectrometry may then fail for the particular case of Quirks and therefore all previous limits may be inoperative.

An interesting curiosity is that if one say $\bar{Q'}$ end of the string is stuck in some rock at the under the ocean or near lake then the partner $Q'$ at the other end is likely to be near the bottom or more generally near the edge of the body of water. It can be stuck there temporarily or permanently depending on the value of $\sigma'$ the string tension and the hardness of the material. Searching in this location may then be more likely to find or to put stronger bounds on the Quirks. 

\section{Subsection(viii) the residual glue-balls of the $SU(N')$}
At the color' confinement phase transition the temperature drops to $T'=T'_{crit}=\Lambda'$ the gluon's form $g'g'$ glue'-balls which we denote by $gb's$. We distinguish the temperature T' in the SU(N') sector from that in the ordinary SM sector (T). These temperatures can be different as the interactions between the sectors are negligible. Indeed as emphasized in above the channeling of all the entropy of the many degrees of freedom into the photons and neutrinos makes the temperature $T$ of the $CMB$ larger than $T'$ :$T\sim {2.4 T}$ so that if the gluon's would not have been confined their present density would be $n(g')= (T'/T)^3 n(\gamma)= 0.07\ n(\gamma) = 30 cm^{-3} $.

The lightest among the $gb's$ is a $J^{P,C}=0^{+,+}$ scalar of mass $m(gb')\sim {7\Lambda'}=7-700 eV$ for $\Lambda'=1-100 eV$. From now on (gb') refer mainly to the $0^{++}$ lightest glue-ball. 

The large ratio $m(gb')/T' _{crit}=7$ suggests that the number density of the $gb's$ remaining after the  confinement phase transition is finished will be lower than the number of the $g'$ gluons  prior to the phase transition at $T'=\Lambda'$. An importance question both for the viability of the Quirk models in the first place and for the technological applications suggested, is how small the ratio $n(gb')/{n(g')}$ really is?

Simplistic arguments using energy conservation would suggest that 
\begin{equation}
n(gb')/{n(g')} = 1/7 
\end{equation}
For $SU(3')$ the phase transition is of first order. Bubbles of the new phase with a $G'_{\mu,\nu}G'^{\mu,\nu}$ condensate and confined blue-balls initially formed in the background of the $g'$s in the unconfined phase grow, leaving eventually island bubbles of the ``wrong vacuum" of the unconfined phase. This may allow for complications and subtle effects of gravity which we do not fully appreciate. 

The fact that the $g'-gb'$ system is strongly interacting and in thermal equilibrium suggests- if we can neglect interactions in some approximation - that both are described by Bose -Einstein distribution. The Boltzmann $\exp{(-E/T')}$ factor implicit in this distribution suggests that the number density $n(gb')$ of the glue ball' is lower than that of the gluons g' forming them by :
\begin{equation}
n(gb')/{n(g')}=\exp{-m(gb')/{T'}}=\exp({-m(gb')}/{\Lambda'})
=\exp{(-7)}\sim{10^{-3}}
\end{equation}
With the original $g's$ number density being half that of a neutrino namely : $n(g')\sim 50 cm^{-3}$ we then have : 
\begin{equation}
n(gb')= 7 cm^{-3}\ or\ n(gb')=5.10^{-2}cm^{-3}
\end{equation}
depending on which estimate among the above two we use. 
Using these two number densities and with $m(gb')=7-700eV$ we find that the energy density of the gb' shortly after the phase transition ranges over   
\begin{equation}
\rho(gb')= 0.35 eV- 5 KeV.cm^{-3}
\cite {equation ? relic density due to reduction at the confining phase transition}
\end {equation}

Over most of this range the density is much smaller than that required if the $gb'$s constituted the dark matter with
$\rho(DM)= 0.2 \rho(critical)= 2.5 KeV cm^{-3}$.

This is important since the $gb'$s do not seem to be acceptable  dark matter.For the assumed $\Lambda'=1-100 eV$ 
and $m(gb')=7-700 eV$, the $gb'-gb'$ elastic scattering cross-section is huge:
\begin{equation}
\sigma_{el}\sim {\pi/{(m(g'))^2}}= (10^{-11}-10^{-15}) cm^2
\end{equation}

Such cross-sections are suggested by the diagram of Fig 2. Just as for $\pi\pi$ scattering with quark annihilation and meson exchanges in both s and t channel we have here gluon annihilation and gb' exchange. The range $r$ of the interaction is then fixed by the lowest mass exchanged- the $0^{++}gb'$: $r=1/{m{(gb')}}$. At the low energy and momentum transfer the coupling, $\alpha'$, is by definition, of order unity with no further suppression of the $\pi.r^2$ geometric cross-section. 

\begin{figure}
\begin{center}
\includegraphics{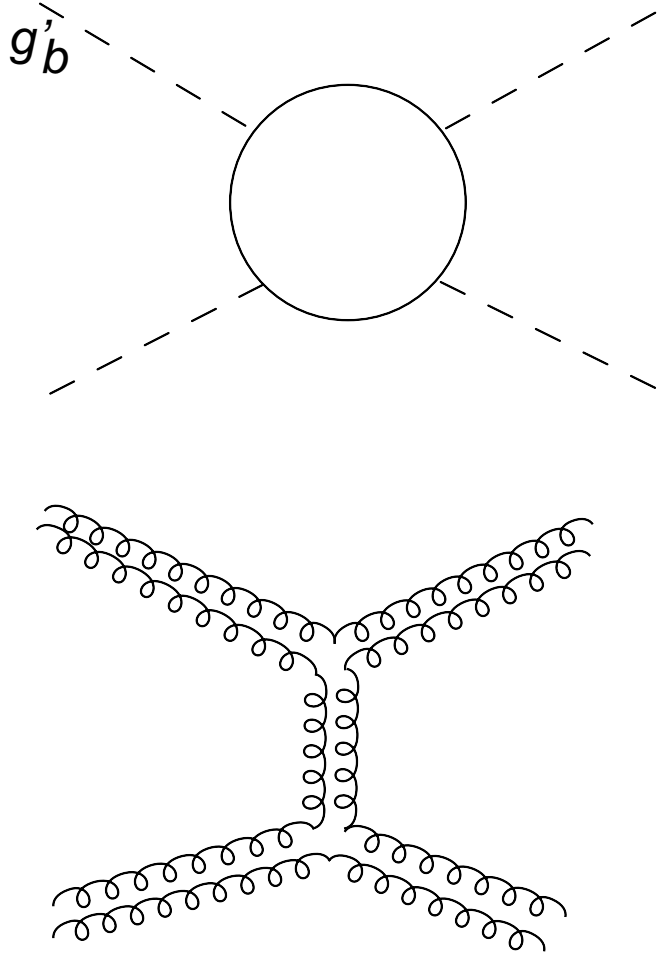}
\caption{An gluon' analog of the Harari-Rosner quark duality diagram for the (2,2) process of scattering of two glue-balls ' ( $gb'+ gb'$). It clearly shows that $g'g'$ i.e the $gb'$ glue-ball states can be exchanged in both the s and t channels}
\end{center}
\end{figure}

An upper bound of $10^{-24}cm^2/{GeV}$ on the ratio $\sigma/m$ of the mutual cross-section of dark matter particles to their mass was inferred from the the bullet cluster and from the the fact that halos can maintain their elliptical form \cite{Feng}. For the $gb'$s with $10^9-10^{13}$ larger cross-sections and masses smaller by a factor of $10^{-6}-10^{-8}$ this bound is violated by $10^{15}-10^{21}$!

Most likely this does disqualify the $gb$'s as a dark matter candidate and limits their present allowed cosmological density by $\sim{1/{10}}$ of the density of standard CDM : $\rho(gb')\le{0.02 \rho_{crit}}=50 e.V$ and the corresponding number density by $n(gb')\le{50 eV/{m(gb')}}$ or $n(gb')\le {(0.07-7)cm^{-3}}$.
 
These bounds could in principle be evaded if the $gb's$ which do attract each other via a non-saturating force due to gb' exchanges condense into ``droplets". The $gb's$ number density inside droplets $n'_{droplet}\sim{m(gb')'^3}=3.10^(16)-3.10^{22}cm^{-3}$ and mass density $\rho'=n'm'=10^{10}-10^{14} Gev/{cm^3}$ or $3.10^{-14}-3.10^{-10} gr/cm^3$ are far bigger than the number and energy densities of CDM in galaxies. 
If sufficiently large $gb's$ condensates form containing most gb's they may render the $gb'$ DM collision-less. 

For $\Lambda'=100 eV$ and corresponding densities giant solar mass ``droplets" would have a radius $R_{droplet}$ bigger than that of the sun by a factor of $(\rho(sun)/{\rho(droplet)})^{1/3}=1.5\times 10^3$ so that $R_{droplet}=10^{14} cm=7 AU$. To account for most of the mass of the galaxy + halo then we need $10^{13}$ droplets within a sphere of volume: $R_{halo}^3 =30\ Kilo-parsecs)^3$. The number density of the droplets $n_{droplet}=10^{23-69}=10^{-56}cm^{-3}$ implies a mean free path for droplet-droplet collisions relevant when two galaxies merge as in the bullet cluster of order:  
$1/{(n(R_{droplet})^2)} \sim 10^{25} cm$ much bigger than the halo size. If most $gb's$ are within such droplets the dark matter becomes effectively collision-less.   

If the structures considered can also evade gravitational lensing bounds then the $gb'$s may be candidates for a very different kind of CDM.\footnote{ it is CDM despite the low mass due to its behaving not as a free-streaming gas but rather as a very sticky fluid} Clearly we need to provide viable scenarios were such giant structures form in the relatively short span between the confinement phase transition at $T\sim{3\Lambda'}$ and the present CMB temperature $T=1/{4000} eV$. Also the droplets may tend to evaporate near the surface because of the self cannibalizing processes of three gb's converting into two gb's which are discussed next and may never achieve these gigantic proportions.

We have not discussed so far the subsequent evolution of the $gb's$ after the confining phase transition which may further decrease their number density beyond the standard dilution by $1/{ R^3}$ with $R$ the scale factor in the Robertson -walker expansion. 

This in turn is related to another important, unique aspect of the $gb's$ that their strong coupling also prevents many particle collision processes from being parametrically suppressed. Unlike asymmetric CDM where the total (co-moving) number of DM particles is fixed the gb's do not carry any conserved quantum numbers. Furthermore at the confining phase transition, the temperature $T'=\Lambda'$  and $\alpha' =1$. The rates of all many- glue- ball processes are then fixed by dimensional considerations :
\begin{equation}
\Gamma'(m,n)\sim{T'\sim{\lambda'}}
\end{equation}
where $\Gamma'(m,n)$ is the rate for $m(gluon')\rightarrow{n (gluon')}$. In particular this applies to  $3gb'\rightarrow{2gb's}$, number changing processes which along with the inverse $(2,3)$ process keeps the system in chemical equilibrium. Along with the $(2,2)$ elastic processes which keep the system in kinetic equilibrium this then guarantees the Bose- Einstein form of the co-moving $gb'$ densities in each $\vec {k}$ mode:
\begin{equation}
dn'/{d\vec{k}}=(\exp{E(k)/T'}-1)^{-1}; E(k)=(k^2+m(gb')^2)^{1/2}
\end{equation} 
The chemical and kinetic thermal equilibrium and the above Bose-Einstein distribution are likely to persist down to very low temperatures and in particular all the way down to the present. To argue for this we focus on the elastic $(2,2)$ processes and assume that the rates for the $(3,2)$ and $(2,3)$ processes are not drastically different.
With $\sigma'(2,2)\sim{1/{\Lambda'^2}}$ the collision rate is $\Gamma(2,2)= n(gb').v'.\sigma'$. Assuming an unattenuated co-moving $gb'$ number density namely:  $n{gb'} \sim {T^3}$ this becomes $\gamma(2,2)\sim T^3/{\Lambda'^2}$ where we used $m(gb')\sim{\Lambda'}$. Comparing this to with the rate of Hubble  expansion which we assume to be controlled by the radiation in the S.M $H\sim T^2/{M_{Planck}}$, we find that the condition for kinetic equilibrium: $\Gamma(2,2)\ge{3H}$ holds down to extremely low temperatures.~\footnote{By dimensional arguments the rate of the $(3,2)$ process is in general also away from the phase transition point: $R(3,2)\sim{n(gb')^2 v'a'^5}$ with $a'=1/{\Lambda'}$. Assuming that the system of $gb'$s is in thermal equilibrium $gb'$s the co-moving number of gb's has the Boltzmann suppression factor $n(gb')\sim{\exp{-m(gb')/T'}}$. The higher power of $n(gb')$ in  the expression above for the rate of $(3,2)$ reaction would seem to suggest that once $T'$, the temperatures of the $gb'$ sector decreases, the $(3,2)$ rate is exponentially smaller than the rate of the inverse $(2,3)$ reaction. This however is \textit{not} the case. The momentum distribution of the $gb'$s has the  equilibrium has the Bose- Einstein form which Boltzmann suppresses also the Kinetic energies of the particles. The center mass energy of the two colliding gb' in the $2\rightarrow3$ process, which clearly is smaller than $E(k_1)+E(k_2)$ has to exceed the $2m(gb')$ threshold at least by $m(gb')$ in order to allow producing the extra $gb'$. This implies that the rate of the $2\rightarrow 3$ is also also suppressed by  $\exp{-m(gb')/T'}$ and detailed balance- a condition for the equilibrium distribution in the first place- with matching rates of the $2\rightarrow 3$ process and its inverse is valid.}

With no chemical potential designed to keep constant the total number of $gb'$s, their number density is exponentially damped by the Boltzmann factor: 
\begin{equation}
n(gb')\sim{\exp{-m(gb')/T'}}. 
\end{equation}
This exponential fall-off is in addition to the dilution in proportion to $1/{R^3}$ of $n(gb')$ where $R$ is the scale factor in the Robertson Walker expanding universe. The latter is implicit in the equation above since the red-shifting of the momenta $k{sim{1/R}}$ scales down the volume element in k space: 
$d(\vec{k})=d^3(\vec{k}\sim 1/{R^3}$.

The conclusion then is that \textit{if} $T'$ decreases in a manner similar to $T$, the temperature of the ordinary photon gas, then the $gb'$s will effectively disappear.
  
The manner by which the co-moving number (and energy density) is evolving here seems similar to the way these are reduced in conventional symmetric WIMP scenarios via annihilation into light particles in ,say, the $SM$ sector. The annihilation effectively stops when the temperature falls down to $T=T_{f.o}$ which is a fraction $1/{x_{f.o}}\sim{1/{20}-1/{30}}$ of the dark matter mass $M(X)$. That happens when the number density of the DM which is suppressed by the Boltzmann factor of
\begin{equation} 
F_{Boltzmann}= exp{(-M(X)/{T_{f.o}})}=\exp{-x_{f.o}} 
\end{equation}
is small enough so that the rate of annihilation $dn_X(anni)/n_X$  becomes smaller than the Hubble expansion rate and a final co-moving density of the dark $X$ ( or $\bar{X}$) particles often expressed in terms the ratio $n(X)/{n(\gamma)}$- ?freezes out? and stays constant.

\begin{figure}
\begin{center}
\includegraphics{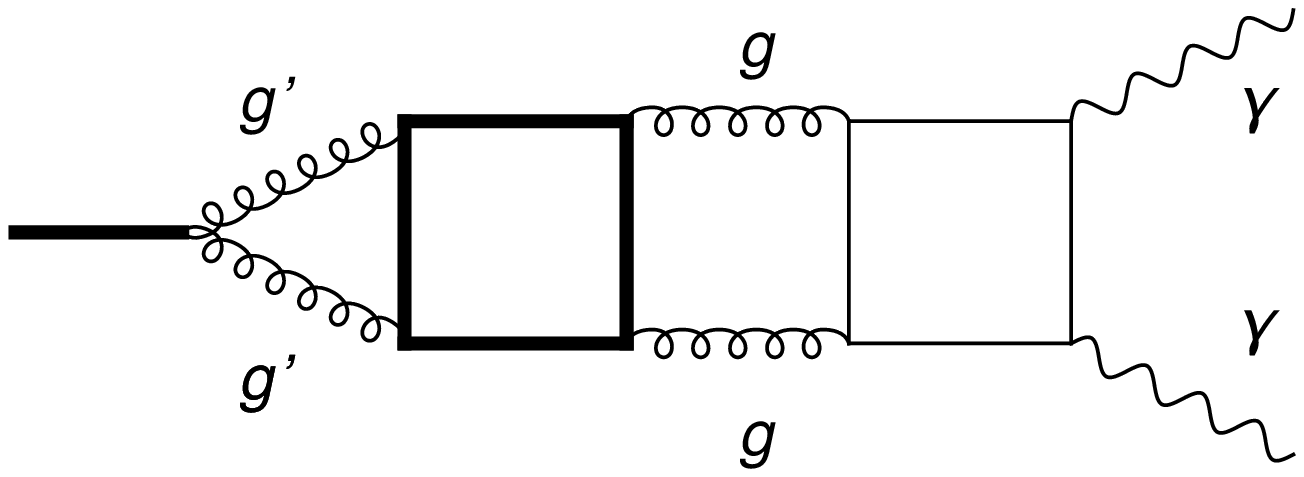}
\end{center}
\end{figure}

\begin{figure}
\begin{center}
\includegraphics{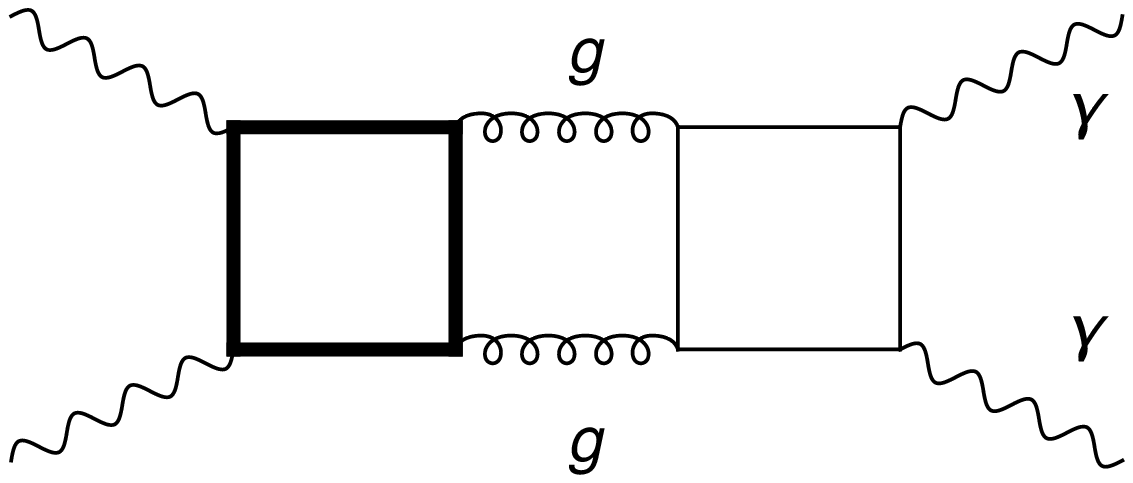}
\caption{The Feynman box diagrams for the coupling of the gb' sector the the SM sector via $g'g'\rightarrow \gamma \gamma$ and to the decay $gb'\rightarrow \gamma \gamma$. The heavy and light lines indicate Quirks or quarks repectively propagating in the two box diagrams. Likewise, the heavy (light) curly lines indicate QCD and SU(N).}
\end{center}
\end{figure}

In the present case the $gb'$ are completely decoupled from the SM and they can neither decay nor annihilate.
Indeed as indicated in Fig(3) FIGURE 3 HREE  $g'g'\rightarrow{gg}$ and the decay $gb'\rightarrow{2\gamma}$ proceed via a Quirk box diagram (followed in the second case by a second quark box diagram). The effective $G'G'GG$ Lagrangian generated by the above Quirk box diagram describing the transitions between $g'g'$ and $gg$ ( with $G$ and $G'$ being short-hand for $G_{\mu,\nu}$ and $G'_{\mu,nu}$ the field strengths of the ordinary color and of the new color') is then suppressed by $(m(gb')/{M(Q')})^4$ and the rates by the extremely small factor of : $(m(gb')/{M(Q')})^8\sim 10^{-80}$. 

The co-moving number of the gb's can then be reduced only via the $(3,2)$ cannibalizing processes where three $gb's$ in the initial state transform into two higher energy gb's in the final state. The degree to which these processes indeed reduce the $gb'$ number and energy density is the issue of interest here. It is equivalent to the question weather the gb's do indeed cool down as the universe expands to temperatures $T' \ll{{m(gb')}}$ so that the Boltzmann factor suppresses their density to almost zero .      

The original paper introducing massive, self- cannibalizing bosons as dark matter candidates \cite{Carlson} suggested that no significant cooling or decimation of the $gb'$s via the $(3,2)$ cannibalization take place.

They argue that since the $gb'$ sector is decoupled from the other sectors it is in thermodynamic terms ``thermally isolated". It should therefore evolve keeping a constant co-moving entropy. In general the entropy density of a non-interacting species of energy $E$ at a temperature $T$ is $s=n.E/T$ with n the number density. For a gas of photons or relativistic 'radiation in general $E$ is proportional to the temperature $T$ and we have the familiar $s(\gamma)=n(\gamma)$. If particles of mass $m$ decay or annihilate and the products are in thermal equilibrium with the ambient gas at a  temperature T then the original entropy $s=nE/T\sim {nm/T}$ transforms into that of m/T as many photons of energy T each and entropy is conserved. \footnote{It is amusing to note that this holds even a black hole of temperature $T_{B.H}$ decaying via Hawking radiation into $M_{B.H.}/{T_{B.H}}$ photons}. 

Here our gas of $gb'$s cannot decay or annihilate into some light degrees of freedom but only to keep self cannibalizing. Once the temperature $T'$ is lower than the rest mass $m(gb')$ by a factor of few the $gb's$ become non-relativistic and the entropy density is $s'\sim {n' m(gb')/{T'}}$. The co-moving entropy $s'R^3$ can stay constant despite the Boltzmann damping:
$n'\sim {\exp {m(gb')/T'}}$ only if the temperature $T'$ of the
$gb'$ sector hardly varies. 

The analysis of \cite{Carlson} then suggested that the co-moving number and energy density would decrease relative to a scenario with a conserved number of CDM particles only by  $1/{\ln(R)}$.

The above logarithmic decrease between the confinement phase transition and the present can be written as 
$\ln(\lambda' /{T_{CMB}})^{-1}= (ln(4.10^5))^{-1}\sim {0.08}$ where $T_{CMB} \sim {1/{4000} eV}$ is the present temperature of the CMB. It further reduces the relic density of the gb's from the estimates of the values given above based only on reductions during the confinement phase transition above to:
\begin{equation}
\begin{split}
n(gb')&=4.10^{-3}cm^{-3} -0.5 cm^{-3} for \Lambda'=100-1 eV\textrm{ and correspondingly:}\\
\rho(gb')&= (0.03 eV -0.35 KeV) cm^{-3}
\end{split}
\end{equation}
Thus the $gb'$ residual density is too small to allow them being  a dark matter candidate a very important point as they would not be an acceptable DM.

For later reference  we recall also the low yet non-negligible cooling down of the gb' s with the FRW expansion due to Carlson, Hall and Machacek:
\begin{equation}
T'/{m(gb')}=1/{3.ln(R/{R_{conf}})} = 1/{3 ln{T_{conf}/{T_{CMB}}}}= 
1/{3 ln{\Lambda'/{{T_{CMB}}}}}= 1/{3ln{10^6}}=1/{40}
\end{equation}
\footnote{The almost constant temperature of a non-relativistic self cannibalizing sector can be avoided if following Ref.~\cite{Hochberg}, we introduce another light sector coupled to the gb's which serves as a sink for entropy. We will not adopt this approach here as we do not wish to change too drastically the Quirk scenario.}

We can avoid all cosmological issues in Quirk models by assuming that the reheat temperatures after the last relevant inflation were much lower than $M(Q')$ and small inflanton gluon' coupling. In this case the Quirks and the gluon's -which generate the $gb's$ were never in thermal equilibrium. This simplistic approach will in particular dispense altogether with the $gb'$ back-ground.
\section{ section III-Quirky strings as lines of communication}
After the above rather long preparation we turn to the main focus of this paper the potential use of the string connecting Quirks for long range communication and in the next section its use for acceleration. Thus assume that we found a $Q'$ and a $\bar{Q'}$ which were pair-produced at the LHC ( or elsewhere) and managed to embed them in two Piezo- electric crystals or some other appropriate solid state structures and or electromagnetic device.  Generating transverse waves traveling along the string by vibrating one crystal and detecting them at the other end makes a communication line with several unique features.

The thickness of the string is $\sim a'=1/{Lambda'}$. This defines the shortest wave that we can send and limits the transmission rate by: $c/\lambda'=10^{16}-10^{18}Sec^{-1}$.  For a variety of reasons we limit the actual rate to much lower values. In particular transmission rates of $\sim{10^7}$ Hertz require only longer carrier waves of  wave-length $\lambda~ 300 meter\sim 10^{10} a'$. For such long waves the flux tube can be viewed as an ideal zero-
thickness string and $(T/{\rho(linear)})^{1/2}=c$ is the propagation velocity.

For finite thickness the propagation speed deviates from  the speed of light $c$ by a wave number $k$ or wavelength 
$\lambda$ dependent small $\epsilon=\sim{(a'/{\lambda)^2}}=
(\Lambda'\lambda)^{-2}= 10^{-18}-10^{-22}$ for $\lambda= 300 meter$ and $\Lambda'$ spanning $1-100$ eV . The dispersion and distortion of the wave packet then limits the range of communication $R$ by$\lambda{\epsilon}^{-1}$ which however is still quite large $L_{com}=10^3-10^7$Parsecs and increasing as $\lambda^3$ with the carrier wavelength.

More complex patterns of communications networks involving many parties can in principle be achieved in the case of $N'=3$ by using the baryonic $Y$ type triple junctions of the $g'$ string as indicated in Fig 4.

\begin{figure}
\begin{center}
\includegraphics{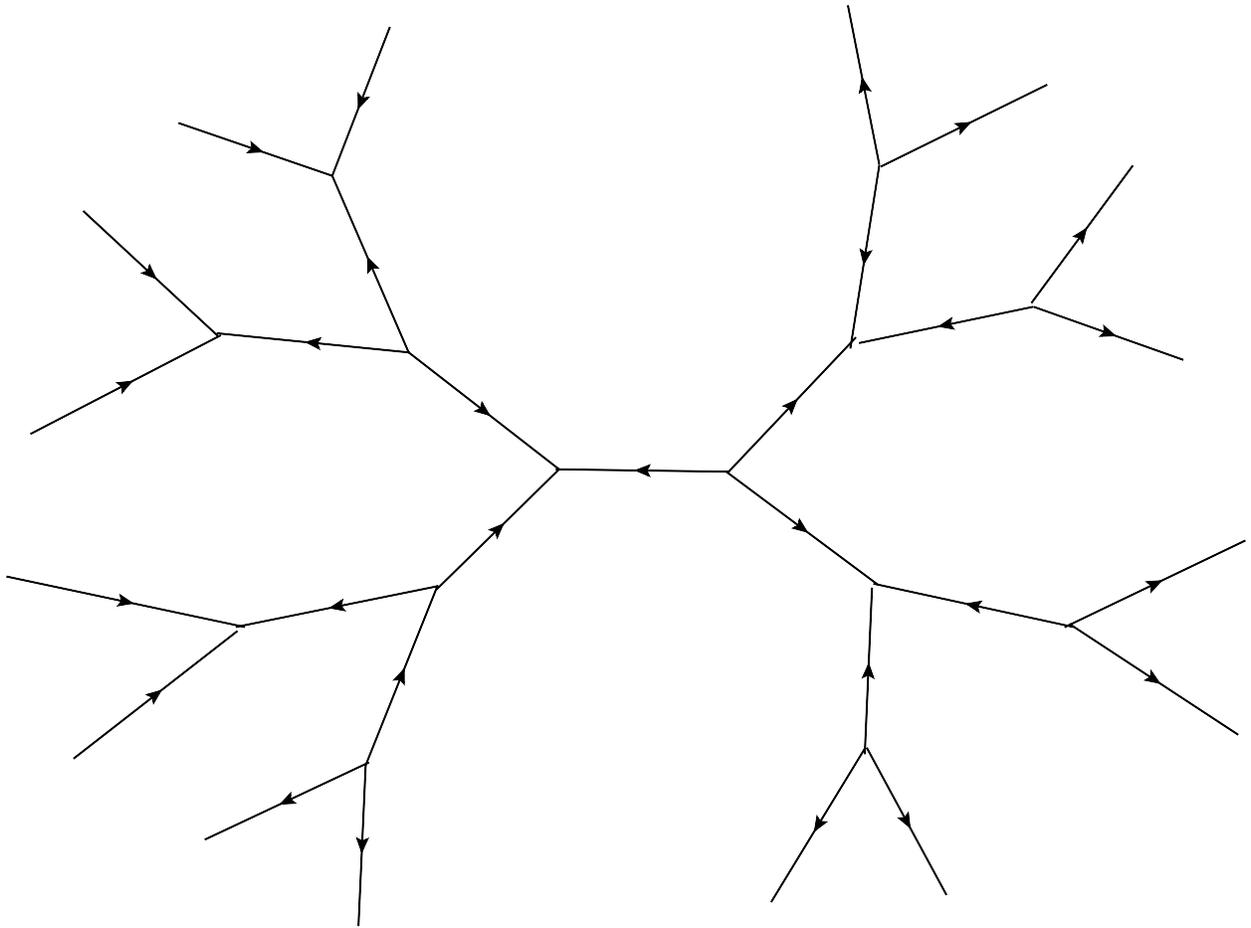}
\caption{A string communication network for $N'=3$.}
\end{center}
\end{figure}

A key technical aspect is the interface between ordinary matter and the $Q'/g'$ system. In sending a sound wave from the piezoelectric crystal to generate the above transverse phonons traveling with the speed of light we may encounter a huge impedance mismatch and the wave will be largely reflected. We will not address the host of such technical issues here but focus on possible more fundamental limitations. 

One question concerns the maximal energy that can be pumped  into our signal or into the dominant carrier wave of wavelength $\lambda$. Let the amplitude of the signal measured by the $\textit{transverse}$ rms velocity be $v_{Tr}$. The energy $\Delta(E)$ stored in a transmitted wave packet or a string bit of length $l=\lambda$ has a kinetic part of $\delta(M).v_{Tr}^2=(1/2)\sigma' \lambda  v_{Tr}^2$ and an equal potential energy. 
In principle we can increase this energy by pushing $v_{T}$ all the way to the speed of light. However limitations of material strength suggest that the maximal velocity is the speed of (transverse)sound: $v(s)\le\sim{\sim{3 Km/{sec}}}= 10^{-5}c$.
Using the above $\lambda=300$ meters and $\Lambda'=100 eV$ we find that
\begin{equation}
\delta(E)= 25 eV [\Lambda'/{100 eV}]^2.\lambda/{300 meter}
\end{equation}      

A string stretched along the straight chord connecting any two points on earth would then afford a noiseless, secure line of communication, faster by up to a $\pi/2$ factor for  antipodal points than ordinary em communication via radio waves proceeding along the earths surface.

For tensions $4\Lambda'^2=(3\times 10^{-7}-3\times 10^{-3})dyne$ we can stretch the string to astronomical distances with minimal effort and negligible energy expenditure. In general the string will be slightly curved due to slow variation of gravity  between various points along the string. Since gravity is a weak force we expect that the string in trying to minimize its energy will follow the Geodesic between the two points where the $Q'$ and $\bar{Q'}$ are located. This would allow mapping the gravitational field on solar system and larger scales with remarkable accuracy.

We note that for higher scale $\Lambda'$ the limits imposed by dispersion on the shortest carrier wave and on the corresponding maximal rate of transmitted information are weaker. However unless $\Lambda'\le{200 ev}$ the corresponding larger string tensions will pull the Quirks at the end of the string out of the hosting crystal.

While our string and the signal traveling along it are immune to any form of em, cosmic ray or other disturbance by SM particle and fields, it is sensitive to the other Quirk strings and to the $gb'$s, two important issues that we will next discuss.

\section{(IIIA) The difficulty with CR's produced strings and it's resolution}

The first issue is again connected with UHE CR's. We need to address effects of their interactions in the inter-stellar medium (ISM) \footnote{I am indebted to Warren Sigel for pointing this out.}. Assume that such a collision produced a $Q' \bar{Q'}$ pair which separate and generate a new string' between them. If the newly produced ``rogue" string collides with our long communication string it can cut it via flux rearrangement. A simplistic model for hadron hadron scattering where the hadrons are modeled by elongated flux tube yields constant cross-sections for inelastic processes generated by flux re arrangements at the intersection region. \cite{Nussinov2}. These are rough approximations for ground state hadrons but become very relevant in the present context. 

We recall that while performing oscillations around the production vertex at the midpoint of the line connecting them in their cm Lorentz frame, Quirks with $0.1-0.5 TeV cm$ kinetic energy separate up to a turning point at a distance of: 
$d_{T.P}= K.E/{\sigma'}=2\times 10^2-10^7 cm$, where $\sigma'=4\Lambda'^2$ with $\Lambda'=100 eV$ and $\Lambda'=1 eV $ respectively.

\begin{figure}
\begin{center}
\includegraphics{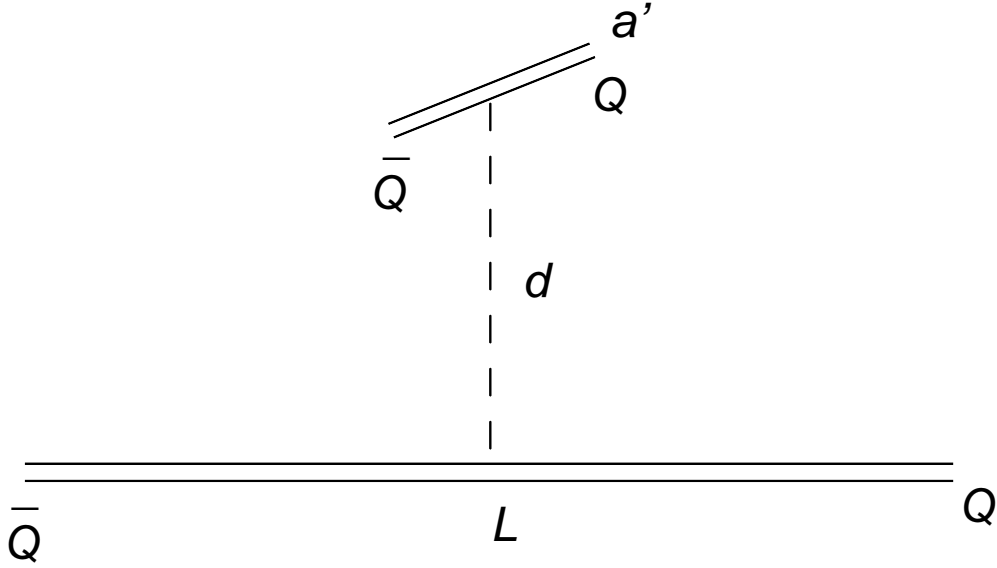}
\caption{A schematic illustration of a short 'Rogue? string produced in the collision of a UHE cosmic ray proton with an ISM Hydrogen, which is  approaching our communication string and threatens to cut it via flux rearrangement.}
\end{center}
\end{figure}

The effective cross-section for the newly formed string bit and our communication string of large length $L$ is $\sigma(s,S)=d_{T.P }.L.f'$  with $f'\sim(1/3)$ a numerical factor reflecting the angle between the strings and the  average length of the rogue string which is smaller than $d_{T,P}$ (see Fig. 5) is very large. The key point is that $\sigma_{s,S}$ is very large. If every collision of a CR with a proton in the ISM resulted in a $Q'\bar{Q'}$ pair then over a time t the number of such collisions that our string would suffer is:
\begin{equation}
N(collisions)= \sigma(string-String)\Phi(CR).t
\end{equation}
with $\Phi(CR)\sim{1.cm^{-2}sec^{-1}}$ the flux of cosmic rays and t the time our communication string is assumed to exist. This time should be mcuh longer than $L/c$ the time required for a string bit or message to travel this far: $t=\tau L/c \tau sec$ with $\tau \gg{{1}}$. From the last equation we find for $\Lambda'=eV$ and $\Lambda'=100 eV$ the number of possible collisions is, respectively:

\begin{equation}
N_{col}= 3.10^{32} L_{18}^2 \tau-3.10^{28}L_{18}^{2}\tau \\
\end{equation}

where $L_{18}$ is the length of the our communication string in units of $10^{18}cm$ or the one light year bench-mark.

Even if just one of these collisions will eventually yield a rogue string which will cut our long string then it becomes useless for very long distance communications.

However we note that

a)For $M(Q')=1 TeV$, only CR's of energies $E\ge{4M(Q')^2/2 m(Nucleon)}=10^7 GeV$ can pair produce the   $Q'\bar{Q'}$ and the integrated flux beyond that point
\begin{equation}
\Phi(UHE(CR)[ E\ge{10^8 GeV} ]\sim 10^{-11}.\Phi(CR) 
\end{equation}
is only a tiny fraction of the total CR flux. 

b) The cross-section for this pair production at these energies is only $\sim {10 femtobarn}$. With a total proton-proton cross-section of $0.1 mb$ the probability of producing the $\bar{Q'}Q'$ pair in a collision is only
\begin {equation}
p\sim 10^{-13}
\end{equation}

Finally a less obvious yet important factor is due to the fact that 

c)The proton number density of ${1/cm^3}$ in the ISM yields a mean free path of the UHE(CR)~\footnote{In passing we note that such high energy cosmic rays are not strongly bent by magnetic fields and are not trapped in the galaxy}: $l_{mfp}= 10^{25}cm$ or travel time of 
\begin{equation}
\l_{travel}/c=3.10^{14} sec.
\end{equation}

The analysis of $Q'\bar{Q'}$ production in vacuum ( see 
subsection (iii) of Sec II above) and the evolution of the resulting short string showed that the $Q'\bar{q}$ and $\bar{Q}'q$ at the ends of the string oscillate, collide and loose energy in each collision. After a time $t_{Q'.F}$(in the lab or galactic frame) they form a Quirkonium and annihilate. 

Specifically we found for $\Lambda'=100 eV$ and
$\Lambda'= 1 eV$ that $t_{Q',F} = 10^{-1} (M(Q')/{TeV}) Sec$ and $3\times 10^3 (M(Q')/{TeV})\ Sec$, respectively.

Thus the production event has to happen within the last $0.1- 10^3$ seconds travel time from our string rather than the $3.10^{14}sec$ implied by the full mean-free path i.e. only in a fraction 
\begin{equation}
3.10^{-15}-3.10^{-12}
\end{equation}
of the cases.
We implicitly assumed that the $ \bar{Q'}Q'$ or rather the mesons $M' \bar{M'}$ will indeed collide at the end of each period and not miss each other in the plane transverse to their main motion in the cm say along the $x$ axis. In complete vacuum this would be guaranteed by conservation of angular momentum in this cm Lorentz frame. The orbital angular momentum initially is $0$ or $1$ and the impact parameter stay essentially equal to $1/{M(Q'}$ so  that also at the end of each cycle the angular momentum stays that small.
The galactic $B$ field of $\sim{10^{-6}}$ Gauss causes a transverse displacement of the TeV quirk after traveling a distance $d_{T.P}$ of $\delta(y)=d_{T.P}^2/{2 R_{gyr}}$. However the displacement collected during the first half of the cycle as the quirk moves say to the right relative to the production point and back is exactly canceled  by the opposite displacement when the Quirk moves to the left and then returns to the original intersection point with the same velocity.
This would not have been the case if the galactic field was varying on a distance scale of $d_{max}$. However the typical persistence scale of the latter is expected to be $\sim{100 parsecs = 1.10^{20}} cm$-many orders of magnitude bigger than $d_{T.P}$

Jointly the above three effects suppress the number of potential cuts by $3.10^{-13-11}.(10^{-15})-10^{-12} \sim{3. 10^{-39}-3.10^{-36}}$. This then reduces the $3.10^{28} L{18}^2$ and $3.10^{32} L_{18}^2$ original would be hits to a probability of a cut to $3.10^{-11}.L_{18}^2$ for the high $\Lambda'=100 eV$ and $3.10^{-4}L_{18}^2$ for the low $\Lambda'=1 eV$. 

Thus we find that for the higher $\Lambda'$ communication strings much longer than a Light year survive for many years.

\section{III-B The effect of the $gb'$ back-ground on long $g'$ string  communication lines.}
In the conclusion of sec II above we found a residual number density of the $gb's$ at present ranging over :
\begin{equation}
n(gb')_{now}=(4.10^{-3}-0.5)cm^{-3} for 
\Lambda'=(100-1)eV 
\end{equation}.

What ``noise" impeding our string communication can be generated by a gb' background characterized by it's mass $m(gb')$, number density $n(gb')$, and velocity $v'$.

The string and the $gb'$ are made of the same $SU(N')$ gauge fields which at the relevant $a'=\Lambda'^{-1}$ scale are strongly interacting. The cross-section that a string bit of length $l$ and thickness $a'\sim {1/{lambda'}}$ presents to a $gb'$ is then $la'$ and during time $t$ $N= n'v'a'lt$ glueballs collide with the string bit. Let us assume that the string bit has length $l=\lambda$, the carrier wavelength encoding one bit of information of the message sent over a distance L. During its travel time,$t_{travel}= L/c$ this bit suffers 
\begin{equation}
N_{col}=n'v' a'\lambda L/c.
\end{equation}
collisions where $a'=2.10^{-7}- 2.10^{-5}cm$ is the width for $\Lambda'=100-1 eV$ . With gb' densities $n(gb)'=(4.10^{-3}-0.5)cm^{-3} $ moving at  a velocity $v'$ this yields a large number of collisions  
\begin{equation}
N_{col}= (2.4 10^{12}-1.5 10^{17}).v'/c =4.10^{11}-3.10^{16} 
\end{equation}
for a distance of $L= 10^{18} cm =Ly$.(Ly=light year). To find $v'/c= (2T'/{(m(gb'}))^{1/2}$ we used the minimal cooling of the $gb's$ in the period since the confinement phase transition due to Carlson et-al quoted at the end of the previous section. 

Can we then transmit information subjected to this barrage of $gb's$ to such a distance?

To answer this question we need to find the impact of any single $gb'$ collision. If the $gb'$ sticks to and becomes part of the string it transfers it's full rest mass namely: $m(gb')c^2=7\Lambda'={(700-7)}eV$ for the $(100-1) eV SU(N')$ scale $\Lambda'$. We find that altogether 
\begin{equation}
\begin{split}
\Delta(W)&= \Phi(gb').\sigma(gb'-string bit].t_{travel}=[n(gb').m(gb')c^2.v'].[\lambda.a'].[L/c]\\
&=4.10^{18}v'/c e.V.\tau =Delta(W)= \Phi(gb').\sigma(gb'-string bit].t_{travel}\sim{10^{18}} eV    
\end{split}
\end{equation} 
would be transferred. 
The three factors in consecutive square [  ] brackets in the above equation correspond to the energy flux of the $gb'$s,to the cross-section for collision of the $gb'$ of radius $a'$with  the string bit of length $\lambda$ and to the travel time across the communication line of length $L$ (for which we use $1$ light year). 

However the sticking of the $gb'$ to the string is unlikely. The point is that the string stretched between the $Q'$ at $R_1$ and the $\bar{Q'}$ at $R_2$ along with it's quantum fluctuation is the true ground state of the complete system subject to the boundary conditions of having the $Q'$ and $\bar{Q'}$ at the above locations. Having the $gb'$ join the string/flux-tube then involves intermediate states of energies higher by the natural scale $\sim{\Lambda'}$ of the theory. These then form effective barriers which the $gb'$, having a kinetic energy $T'\le{\Lambda'}$ can tunnel through but suffer an 
$\exp{-2\Lambda'/{T'}}$ suppression. Using
$T'/{m(gb')}=1/{40}$ estimated above (see the  estimate of v' and the last equation in section II.) we find a suppression of  $\exp{(-80/7)}\sim 10^{-5}$. 
This then would reduce the previous estimate of the energy transferred to the relevant string bit of length $\lambda$ down to $\Delta(W)$ of $10^{13} e.V.$  

To visualize the above argument we can utilize the string theoretic picture where the glue-ball is a small closed string or flux tube turned into a torus.
\footnote {This picture nicely explains the $\sim{\pi}$ ratio between lightest $\bar{q}q$ mesons viewed as open strings and the lowest gb' small closed string}

\begin{figure}
\begin{center}
\includegraphics{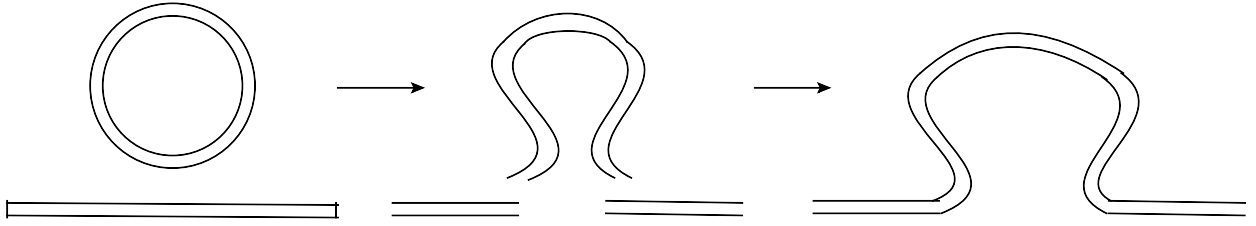}
\caption{A pictorial illustration of the topological and ensuing energy barrier which suppresses the process of a tiny closed string ( our glueball $gb'$) sticking to the long communication string.}
\end{center}
\end{figure}

As illustrated in Fig (6) the capturing of the $gb'$ requires that both the long linear string and the small closed string representing the $gb'$ will open and reconnect. Clearly this entails intermediate stages where the energy of the system is higher than the sum of the original energies of the string and the glue-ball($=m(gb')$) by the typical scale of the theory $\Lambda'$.
This issue has been encountered earlier \cite{Jacoby} in connection with the inverse process. We wondered then if the string between the $Q'$ and $\bar{Q'}$ quickly shortens by evaporating $gb'$s at a uniform rate all along its length a possibility which is ruled out by the very same argument as above.

Let us next consider elastic $gb'$ scattering off our string bit of length $l=\lambda$. Only the fraction $f$ of the kinetic energy $m(gb')v'^2$ of the scattered glue-ball which is transferred into string modes with the carrier frequency $\omega_0=\lambda/c$ is relevant. Higher frequencies contribute ``white noise" readily filtered by Fourier transforming and cannot destroy the signal.  
To estimate the fraction $f$ we note that the range $a'$ of the $g'$ forces in the present confining phase, which is also the thickness of the $g'$ string, yields an effective time duration of the $gb'$- string' encounter of order: $\delta(t)= a'/v'$. The key point is an impulsive and local pinching of the string is not likely to produce perturbations (i.e waves) on the scale $\lambda$ and with the frequency $\omega(0) = c/\lambda$. Indeed Fourier transforming the gb' -string impulsive interaction of duration $\delta(t)$ and squaring the resulting amplitude we find :   
\begin{equation} 
f= (\omega(0).\delta(t))^{-2}= (a'/{v'}).(c/{\lambda})^2
\end{equation}
The above assumed a sharp square well type form for the pulse .The fall off of the interaction with distance ( and with time) is much more gradual which considerably strengthens the present argument. Thus had it been a Yukawa type interaction we would have a quadratic fall-off of the amplitude and a much faster quartic fall off: $ f= (a'/{\lambda})^4$.
Inserting even the more conservative damping factor into our earlier estimate of the ?Noise? energy $\Delta(W)$ and substituting the kinetic energy $m(gb')v'^2$ of the $gb'$,  instead of $m(gb').c^2$ for the energy transported in an individual $gb'$ collision we find that the relevant noise energy at the frequency considered is:
\begin{equation}
\Delta(W)_{\omega(0)}= [n(gb').m(gb')v'^3].[\lambda.a'].[L/c].[(a'/{\lambda})^2 .(c/v')^2]=7.(n(gb')a'^2.L) (v'/c)
(a' \Lambda')/{\lambda}    
\end{equation} 
The factor f was added as the forth factor in square [..] bracket on the second line and dimensionless quantities  ( including $\Lambda'.a'=1$) in the last line.

The issue is then weather this noise energy $\Delta(W)$ exceeds the actual energy carried in the string bit of length $\lambda$ which was estimated above:  
$\Delta(E)= \lambda.4\Lambda'^2 u^2$ with $u\sim 10^{-5}c$ the sound velocity. The ratio  $r=\Delta(W)/{\Delta(E)}$ then is:
\begin {equation}
r=7n(gb')(a'/{\lambda})^2 (L/{\Lambda'})^2.v'/{u^2}=
10^{-6}n(gb')(\Lambda'/{100 eV})^{-4}.(L/{10^{18}cm}).(\lambda/{3.10^3 cm})^2.
\end{equation} 
Where we have displayed the dependence on the relevant Quirk model parameter $\Lambda'$ and on the controllable parameters $L$ the length of the communication line and $\lambda$ the carrier wavelength that we chose to correspond to $\sim 10^6$ Hertz carrier frequency.    
The dramatic $\Lambda'^4$ increase of signal to background found here confirms a recurring theme that the applications considered including the acceleration discussed in the next section improve the higher the new scale $\Lambda'$ is. Unfortunately this parameter just like the very existence of Quirks in the first instance are fixed by``nature" and is not something that we can change. Also the limited strength of known materials does not allow benefiting from Quirk models with string tension much exceeding   
$\sigma'= 4.10^4 eV^2 = 20 eV/{Angstrom}= Ry/{r_{Bohr}}$ 
which corresponds to $\Lambda'=100 eV$
Using our estimate $n(gb')=4.10^{-1} -4 10^{-3}$ for the residual $gb'$ number density we then find that unless $\Lambda'$ is smaller than $3 eV$ we can transmit information to Light year distances at a rate of about $10^6$ bits per second. Using $10^3$ times longer wave-lengths $\lambda$, which decreases the rate to only $10^{3}$ bits per second would allow increasing the range of communication by a factor of a million.  Also just as for ordinary radio cellular communications the range can be increased by using amplifying relay stations at the junctures of consecutive Quirk strings.

``Modest" $L=10^{15}cm =$ solar system size communication goals could be achieved even for a $gb'$ background with the same number density as that of the original gluon's, the latter being less than half of that of a single neutrino: $n(gb')=1/2 n(neutrino)\sim{50 cm^{-3}}$ 

\section{Sec IV. Can Quirky strings accelerate particles to very high energy?}
The UHE cosmic rays CR's at $\sim{10^{11}} GeV$ are the highest energy ``almost" elementary particles known. Used as projectile on fixed target protons they can generate cm energy $W=(2m(N).E_{CR})^{1/2}$ of up to $500 TeV$, forty times larger than that of the LHC. If $SU(N')$ strings of the form discussed above with string tensions  $\sigma'=(4-4.10^4)^(eV)^2 $ exist then they can accelerate the Quirks and the embedding nuclei at the end of the string to far higher CM energies.

Assume that the grains embedding the Quirks at the ends of the string were separated to a distance $L_{S}$ ranging between $10^{9}cm$, earth's diameter size, to solar system $10^{15}cm $ sizes. The energy stored in the string is then  $\sim{10^2-10^6 and 10^{8}-10^{12} TeV}$ respectively where the inner range corresponds to $\Lambda'=1 -100 eV $ or a string tension $\sigma'=4-4.10^4 eV^2$.

\begin{figure}
\begin{center}
\includegraphics{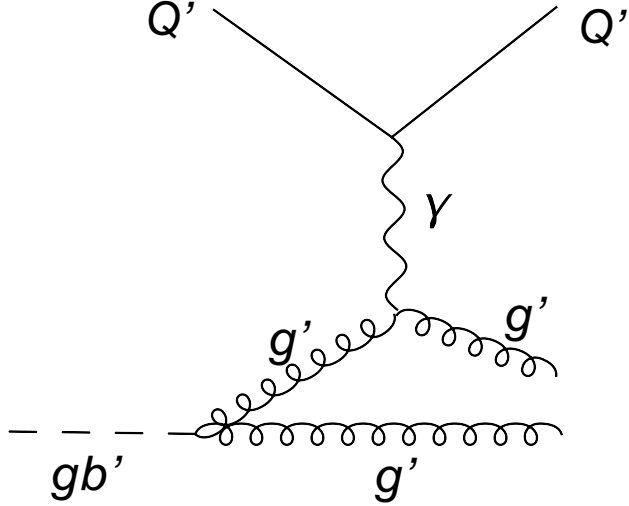}
\caption{The diagram for the Quirk glue- ball scattering, which is some-what reminiscent of that for the scattering of electrons on protons. Here the gb' glue-ball made of two $g'$ gluons is the analog of the proton target made of  quarks, the $Q'$ projectile is the analog of the electron, and the exchange weakly coupled g' is the analog of the virtual photon.
}
\end{center}
\end{figure}

We evaporate via an intense laser pulse, or otherwise, the grains in  which the $Q',\bar{Q'}$ are embedded. \footnote{If some small grains still stick to the heavy mesons M and $\bar{M}$ at the end of the string they will be anyway evaporated early on by the impact of the CMB photons. The $\sim{3 Kelvin}$ temperature of the latter is blue shifted by a Lorentz factor $\gamma$ in the grains rest frame and once $\gamma\le{10^3}$ will melt and vaporize it away.}
Most of the energy gained by the acceleration will then be available for the eventual collision of the $\bar{Q'}Q'$. A portion of $M(A,Z)/{M(Q'}\sim 0.1-0.01$ of the energy  will be available for collisions of $Q'$ and the nucleus $(A,Z$ attached to the $\bar{Q'}$ at the other end and $M(A,Z)^2/{M(Q')^2}\sim{10^{-2}-10^{-4}}$ for nucleus nucleus collision.
Since the $Q'$ carry color $\bar{Q'}Q'$ collisions will be very effective in producing more $Q'\bar{Q'}$ pairs and $\textit{any}$ other new particle particularly if they carry color (or color') 

By moving the quirks to two symmetric points equidistant, in the appropriate frame and metric from an LHC size special detector we can have the collisions occur inside the better controlled environment of such detectors.

An earlier work $\cite{Casher2}$ argued that no device can accelerate a single ``elementary particle" to energies exceeding $M_{Planck}$. No single overriding argument was invoked but many putative Planck accelerators were shown to fail in the context of existing physical laws and materials and our astrophysical milieu
\footnote{From some prospective this may be fortunate as such collisions could produce mini-black hole which if they are stable enough against Hawking radiation, may grow and 'eat up? earth or more. The idea and ensuing concern that even at the LHC some much lighter mini-black holes associated with a much lower Planck scale of a theory with  extra dimensions, can be produced and have catastrophic consequences has been debunked as such objects would have been produced long ago by UHE CR's of energies exceeding that of LHC. Since cosmic ray energies are twenty seven orders of magnitude short of producing true Planck mass particles this argument is not applicable here.}

Huge advancement and insights would be gained if we could close just part of the gap between LHC and Planck energies. It is still interesting to find if basic physics constraints the maximal energy accessible via our ``Quirk accelerators".
Eventually space travel or other techniques may reach the  separation distance of $Ly=10^{18}cm$ and formally Quirks could then be accelerated to (super) Planck energies.

Once $\gamma=E/{M(Q')}$ of the accelerated $Q'$s exceeds $\sim{10^{11}}$ a GZK effect kicks in and the u and d quarks attached to the quirks will start colliding with the CMB photons. As discussed early in Sec II the energy lost in each collision will be a small fraction $f\sim{GeV/{M(Q')}}=10^{-4}- 10^{-4}$ of the running Quirk energy $E$. The $\sim {0.1}$ fraction of energy lost in collisions of UHE cosmic ray protons with the CMB limits via the standard GZK effect the range of UHE cosmic rays of energies exceeding the $10^{11}(GeV)$ cut-off to about $100$ Mega-Parsec. The GZK range in the present case is, due to the small f above, $\sim{10^3}$ times larger than the above $\sim {100 Megaparsec}$ range makes it irrelevant in the present case in particular since supper Planck energies can in principle be achieved over much shorter $\sim 30$ Parsec acceleration paths. 

The $gb'$s background seems to present a more serious problem: The target masses of $7-700eV$ are $10^4-10^6$ times higher than the energies of the photons in the  CMB and the threshold for inelastic scattering involving here just production of extra $g'$s is much lower than the $\sim 200 MeV$ threshold in the case of the  Compton scattering on nucleons.
The relevant process is analogous to inelastic scattering of electrons off protons, where the role of the energetic electron is played by the Quirk, the exchanged  photon is replaced by the exchanged $g'$ in the diagram of fig 4 and it probes here the $gb'$ which is made of the two gluon's. The cross-section $\sim{\alpha'^2/{m(gb'}^2}$ is quite large and for a $gb'$ density $n(gb')$ in $cm^{-3}$ units the mfp for a $Q'-gb'$ collision is 
$l_{mfp}=2.5 10^9/m(gb')/ev^2 /{n(gb')\alpha'^2}$ where $2.5 10^9$ is the $(eV.cm)^2$ conversion factor. In each collision only a fraction $m(gb')/{2M(Q')}$ is transferred so that many collisions:
$N_{co}\sim{2 M(Q')/{m(gb')}}$ are needed to slow it down and the path length required is correspondingly longer : $L(stop)= N_{co}l_{mfp}$ Using  $m(gb')= 70 eV$ we find  $N_{co}=10^{11}$.  We find a stopping distance: $L(stop)= 5 10^{23} cm/{n(gb')\alpha'^2}$ which for $\alpha'=0.3$ and $n(gb')\le{1}$ is far larger than the $10^{22} cm$. On accelerating along a $10^4$ times shorter distance of $5.10^{18}cm$  a  string' with tension of $\sigma'= 4.10^4 eV^2$ would yield Planck energy Quirks. The unhindered Quirks will keep oscillating as we discussed at length at various points above for some $10^4$ times before slowing via mutual collisions, binding into a Quirkonium and annihilating at rest.

There are clear drawbacks of our accelerator. To begin with we have only one pair of Quirks that we may collide as compared with the $10^{17}$ collisions in the LHC high intensity proton-proton machine. Furthermore the much bigger cross-section for the mutual scattering the attached $q$ and $\bar{q}$ may prevent us from having in most cases  even one high energy collision of the quirks before they will slow down and stop even in complete vacuum. 

Other hurdles impeding our Quirk accelerators include the following: Unless we shield the galactic $B$ fields all along the  $L=10^{18}cm$ acceleration path the transverse deflections of the Q' and bar Q': $\delta(y) = L^2/{R_{Gyr}} = 10^9 cm$  due to the galactic B fields will now be much larger then in the case of $L'=10^4 cm$ long rogue strings studied at the conclusion of sec III. Due to the much higher energy the Gyro radius here is $R/R'=M_{Planck}/{m(Q')}\sim {10^{15}}$ times longer but $\delta(y)= L^2/R_{Gyr}$ is $10^{17}$ time larger in the present case. Also unlike the previous case the galactic $B$ fields vary slightly over $L=10^{18}$ distances and the complete cancellation over a period encountered there no longer applies. Thus the accelerated particles may miss each other. In principle we could try and correct for it. Such corrections are impossible for the linear accelerator considered in \cite{Casher1}. The reason being that as E(p) tends to $M_{Planck}$ the velocity of the proton: $\beta= 1- [m(p)/{M(Planck)}]^2$ approaches the speed of light to within  $10^{-38}$. This exceeds the speed of radio or optical photons in the galaxy and undercuts efforts to signal and correct for any deviation. The situation in the present case is much better in this respect:
Here $\delta(\beta)=(m(Q')/{M_{Planck}})^2$ is $10^6-10^8$ larger. Furthermore the repeating almost periodic motion of the Quirks offers us many $10^3-10^4$ opportunities to correct the trajectories so as to achieve eventual intersection inside the desired volume of the detector. 

\section{Section V Quirks at the LHC}
Since the signatures of quirks produced at the LHC have already been discussed in some detail we will only briefly touch on this issue. Assuming $\Lambda'=30-100 eV$ which was more favorable for the applications suggested, the separation between a produced $\bar{Q'}$ and $Q'$ of $TeV$ mass until their turning point is on average 
$d_{TP}=(10^2-10^3)cm$. The bizarre translational+ relative back and forth Yo-Yo like motion makes it difficult to identify the Quirk, or rather the heavy $\bar{Q'}q=M$ or $\bar{q}Q'= \bar{M}$ mesons as fractionally charged heavy particles which reach the muon detectors and which also produce some pions along the way. Without a dedicated analysis it is not clear what are the present bounds on the production rates of such particles. 
  
Identifying the rare events in which $\bar{Q}' Q'$ may have been produced among the many billions of LHC triggers seems an impossible task and in any event provides only limited  information on where the Quirks have stopped.
After the LHC operation is completed and a waiting period  required for lowering the radioactivity in the nearby ground to acceptable levels one could in principle start looking for the Quirks there.  
How can we find in the $\sim{0.1 Km}^3$ of crust material surrounding the LHC detectors even one of the say $10^{2}$ Quirks that have been produced over the whole LHC operation and which stopped in this near-by volume?. We cannot break it down to single atoms and put them through isotope separators until Quirks are identified. 
  
The pull on the $Q'$s embedded in grains by a $10^{-1} dyne$ string tension appropriate for the optimal $\Lambda'=100 eV$ overcomes gravity for $M_{grain} = 4.10^{-6} gr$. This suggests  a more efficient search where deviant grains containing a $Q'$ or $\bar{Q'}$ are spotted by their weird behavior of not falling straight down. The number $\sim{10^{17}}$ of such grains in that volume is still huge but each grain contains some $\sim{10^{17}}$ atoms! Once deviant grains are identified we proceed by breaking them to smaller ones and quickly converging on one end of the string attached to a $Q'$ or $\bar{Q'}$.
\newline
Another issue is transporting the other end of the communication line, the $Q'$ (or $\bar{Q'}$) to it's far out designed location- which may take very long time if no new method, such as repeated sling shooting via the above quirk acceleration, is found.
  
\section{Sec VI Further Comments on Quirks and Quirk models.}
We do not address here the source of the large Quirk mass
$M(Q')\ge{TeV}$ beyond the following incomplete sketch.
  
Starting with a common $SU(3)_c$ and $SU(2')$ small coupling at a common high energy scale naturally predicts a tiny ratio of $\Lambda'/{\Lambda(QCD)\le{10^{-6}}}$ due to the slower running of the $SU(2)$ coupling as compared with that of $SU(3)$. 
  
Let us then assume that the Quirks are to start with mass-less but carry yet another $SU(4'')$ gauge interaction. With $\alpha''=alpha_c=\alpha''$ at a common high scale the faster R.G running of $SU(4)$ generates a scale $\Lambda''$ much larger than $\Lambda(QCD)$. $SU(4'')$ then confines the Quirks at this higher $\Lambda''$ scale and like in QCD generates also a spontaneous chiral symmetry breakdown and effective Quirk masses. The constituent light quark masses $m(q)\sim{\lambda(QCD)}$ suggest that also the spontaneously generate effective Quirk masses $M(Q')\sim {\Lambda''}$. These confined Fermion'' cannot be the Quirks of the above Quirk model as they are confined at a very short, $1/{\Lambda''}$, distance. 

However the SXSB associated with generating Quirk masses leads also to pseudo-scalar Nambu Goldstone bosons (PNGB). Again following the known pattern in QCD only those bosons which are not complete singlets namely those which carry also  color and or color ' remain light. They do however get radiative masses due to color and or color' interaction of order $\alpha(QCD)\Lambda''$ or $\alpha'.\Lambda''$. These spin-less particles, rather than the Original confined Fermions $F'$s, which play the role of the Quirks here. Because of their small size they can be viewed as elementary on the scales we are interested in. Also if $\Lambda''/{\Lambda(QCD)} =10^5 i.e. \Lambda''=20 TeV$ then radiative masses of order few $TeV$ naturally result for them.   
  
Many such PNGB particles exist and the ones we are interested in, those which carry \textit{both} ordinary color and color' are not the lightest ones. The PNGB's which carry just $SU(2')$ quantum numbers will not be produced in the LHC and those which carry only $SU(3)_c$ will be produced but with no long strings attached will have no distinctive characteristics. Unfortunately we need to avoid  decays of our heavier bosonic Quirks which carry both $QCD$ color and color' into lighter ones carrying just one of the two colors. While this can be achieved it complicates this otherwise simple and suggestive picture.   
  
We have assumed that the Quirks are matter-anti-matter (charge conjugation) symmetric. Absent a framework explaining their mass it would be even more futile to try and estimate a possibly asymmetry in the Quirk sector (which is clearly the case recalling that we do not have a  compelling explanation of our baryon-anti-baryon asymmetry)   
  
Let us still assume that some small fraction $\eta_{Q'}$ of the quirks consistent with them being (part of) the CDM remained after the symmetric part annihilated. As described in II above ,in the $N'c=3$ case, these will be mainly in the form of tightly bound $Q'^3$ ordinary color singlet baryons  which are however $SU(3')$ deca-plets. Early on these baryons can form a new type of (dark?) matter.

As noted earlier their hidden color may still generate cross-sections which will allow detecting them in direct DM searches. There is also another very interesting possibility. Many such $Q'^3$ baryons can bind together via the color' interactions forming analogs of ordinary solids. The density of the latter is fixed by: $\rho_{matter}\sim{m(N)}\alpha(em)^3/{m_e^3} $ to be $\sim{gr/{cm^3}}$ . For the putative quirk matter we have $\rho_{Quirk}\sim{\alpha'^3}\sim {[M(Q')^4]}=10^{22}gr/{cm^3}$  even for $\alpha'\sim \alpha(em)=1/{137}$ and $M(Q')=TeV $
  
Clearly even small grains of the new Quirk matter would have a huge mass. It is impractical to produce large quantities of Quirks at LHC and very difficult to detect quirky nuggets. Still we note that having novel supper-dense and supper-strong materials may opens new technologies, particularly if they could be integrated with ordinary matter. 

Thus the Quirk model considered offers two new types of matter  one of which, the $gb'$s droplets, having density of $\Lambda'^4$ which is about  $10^{4}$ times more dilute than that of the above quirk grains. 
  
\section{Sec VII  Summary and conclusions}

We have no evidence nor any good motivation for Quirks in general and for Quirks with the $\Lambda'$ in the range required for communication and or acceleration in particular. The above  amusing exercise of discussing this possibility involved many physics issues. The very fact that particles of potential technological use can exist and be produced by the LHC (or similar higher energy machines) is in itself of great importance.
  
Indeed for the last several decades it was taken for granted that \textit{no} technological implications will trace back directly to particles discovered in HE accelerators.
Still this research which is becoming almost prohibitively expensive is vigorously pursued by several international collaborations. The returns beside various spin-offs are in the realm of extending our fundamental knowledge of nature.
This attitude was most nicely expressed by Robert Wilson who, when asked by a congressman, what contribution the then planned Fermi-lab accelerators, would make to the defense of the US, answered: ``none, but it will make it more worth defending!"
  
As a member of the high energy scientific community which greatly benefited from this attitude and by conviction I espouse this attitude. Still refuting the complete impossibility of benefiting directly from possible future discoveries along the above or similar lines is worthwhile.
  
\section{Acknowledgemnts:}
I would like to thank N. Arkani-Hamed, P. Bedaque, T.D. Cohen, D. Curtin, L. Fitz-patrick, J.Gates, P. Ginsparg, I. Goldman, Y. Grossman, D. Karzeev. J.Rosner, R. Shrock, W. Siegel, R. Sundrum and S. Wallace for comments and/or encouragement and Maryland Center for Fundamental Physics for its hospitality and in particular S. Eno and T. Jacobson. I should also thank anonymous referees who pointed out some shortcoming of my earlier paper with C. Jacoby and encouraged to think more carefully about the model. Finally I should thank members of the extra-solar planet search group of Prof Tzvi Mazeh at Tel-Aviv university for discussions which indirectly inspired this work during discussions of extraterrestrial intelligent societies. 

\footnote{Extra-terrestrial, intelligent life has been a long standing conjecture and the efforts to detect signals from advanced technological societies on extra-solar planets motivated the SETI project. Various arguments suggest that this is prone to fail, even if highly advanced extra-solar societies do exist. In particular the extremely advanced pattern of communication and the actual physical carriers of information used by these societies may be very different from ours. Our efforts to detect signals sent by them via radio waves or optical photons may then be as futile as those of Amoebas trying to listen to and appreciate a piece of classical music (Assuming that some-one cared to play it for them...) Clearly the existence of advanced extra-solar societies which are attempting to communicate with us and of technologically useful quirks are logically distinct possibilities each with unknown and small probability. Yet if both are realized then most likely the extra-terrestrials (E.T's) have already discovered Quirks and use them as means of communication.}


\begin{thebibliography}{25}

\bibitem{de-Rujula} A. De-Rujula, S.L.Glashow and U. Sarid Nucl Phys B 333, 173 (1990).

\bibitem{De-Rujula2} A. De-Rujula, S.L.Glashow, R. R. Wilson and G.Charpak Phys. Rept. 99, 341 (1983).

\bibitem{Lowey} A. Lowey, S. Nussinov and S.L.Glashow arXiv eep-ph 1407 4415 .

\bibitem{Drukier} A. Drukier and S. Nussinov Phys.Rev. Lett. 49 ,102 (1982).

\bibitem{Kang1} J. Kang, M. luthi and S. Nasri JHEP 0809,086(2008) arXiV ; hep-ph/06 11322.

\bibitem{Kang2} J. Kang and M. Luthi arXiv hep-ph 0805.4642

\bibitem{Jacoby} C. Jacoby and S. Nussinov arXiv hep-ph 07.12.268

\bibitem{Nussinov1} S. Nussinov and C. Jacoby arXiv hep-ph 09 07 4932

\bibitem{morningstar} L.J. Morningstar and M. Peardon Phys.Rev D 56 4043(1997) and ibid Phys.Rev D 60 03 4509 (1999)
\bibitem{Mitov} M. Czakon, A. Fiedler and A. Mitov hepph
13036254

\bibitem {Schwinger} J. Schwinger Phys Rev 82.b664 (1951)

\bibitem {Casher1} A. Casher, H. Neuberger and S. Nussinov Phys rev D 20 (1979).

\bibitem{Nussinov2} S. Nussinov, Phys. Rev. D \textbf{50} 3167 (1994).

\bibitem{Z. Nussinov} Z. Nussinov and S. Nussinov arXiv condmat 0409094

\bibitem{N. A. Hamed1} N. Arkani-Hamed and S. Dimopoulos JHEP o6 073(2005)

\bibitem{Feng} J. L. Feng, M. Kaplinghat and Hai-Bo Yu Phys. Rev. Lett 104 15 1301 (2010)

\bibitem{Carlson} E.D. Carlson, M. E. machacek and J. L. Hall ? Self interacting dark matter? HUTP-91-A066

\bibitem{Hochberg} Y. Hochberg, E. Kuflik, T. Volansky and J.G. Walker arXiv hepph 1402.5143

\bibitem{Casher2} A Casher and S. Nussinov hepth 9709127

\bibitem{N. A. Hamed2} N. Arkani-hamed, C. Cheung, J. Kaplan and T. Vachaspati

\bibitem{Hemick} M. Themick et al, Phys. Rev. D \textbf{41} 2074 (1990). 

\end{thebibliography}
\end{document}